\documentstyle[prl,aps,multicol,epsf]{revtex}
\draft 

\begin{document}

\title{Atomic scale engines: Cars and wheels}

\author{Markus~Porto, Michael~Urbakh, and Joseph~Klafter}

\address{School~of~Chemistry, Tel~Aviv~University, 69978~Tel~Aviv, Israel}

\date{December 16, 1999}

\maketitle

\begin{abstract}
We introduce a new approach to build microscopic engines on the atomic scale
that move translationally or rotationally and can perform useful functions such
as pulling of a cargo. Characteristic of these engines is the possibility to
determine {\it dynamically} the directionality of the motion. The approach is
based on the transformation of the fed energy to directed motion through a
dynamical competition between the intrinsic lengths of the moving object and
the supporting carrier.
\end{abstract}

\vspace*{-2mm}
\vspace*{-4mm}

\begin{multicols}{2}
The handling of single atoms and molecules has become widespread in science
\cite{Science:1999}, but the challenge still remains to further `tame' them and
make single molecules perform useful functions. Nanoscale technology has been
predicted almost 40~years ago \cite{Feyman:1960/61}, but inspite of a growing
interest in atomic scale engines, such as biological motors
\cite{Howard:1997,Huxley+Howard:1998}, ratchet systems
\cite{Howard:1997,Astumian:1997,Juelicher/Ajdari/Prost:1997,%
Thomas/Thornhill:1998}, molecular rotors
\cite{Balzani/Gomez-Lopez/Stoddard:1998,Sauvage:1998+Armaroli/etal:1999,%
Gimzewski/etal:1998,Kelly/DeSilva/Silva:1999+Koumura/etal:1999}, and molecular
machinery in general \cite{Drexler:1992}, a real breakthrough concerning the
construction of a man-made nanoscale counterpart of the `steam engine' has not
occured yet. This has mainly been due to the fact that we still miss the
crucial link of how to transform energy to directed motion on this scale.

In this Letter we propose possible basic principles of such an engine. The main
advantages of this novel approach are: (a)~the same concept applies for both
translational and rotational motions, (b)~the directionality of motion is
determined {\it dynamically} and does not require spatial asymmetry of the
moving object or of the supporting carrier, (c)~the velocity obtained can be
varied over a wide range, independent of the direction, and (d)~the engine is
powerful enough to allow for the transportation of a cargo.

The proposed engine consists in general of two parts: the supporting carrier
and the moving object. Achieving motion of the engine is based on dynamical
competition between the two intrinsic lengths of the carrier and the object.
This competition is used to transform initially fed energy to directed motion.
To exemplify the concept, we use below a {\it simple} model system of a chain
in a periodic potential, namely a Frenkel-Kontorova type model
\cite{Frenkel/Kontorova:1938}. But we would like to emphasize that this choice
as example is solely motivated by the simplicity of the model rather than by
experimental requirements. In particular, the particles are not meant to be
single atoms and the springs are not meant to be single chemical bonds. The
sole purpose of the model system is to address in a simple manner the following
questions: (a)~What is the minimal size of the engine? (b)~How are the
direction and velocity of the motion determined? (c)~How powerful is the
engine? The important questions of possible physical realizations will be
addressed towards the end of the Letter.

\begin{figure}[h]
\vspace*{-3mm}
\def\epsfsize#1#2{0.775#1}
\noindent\hfill\epsfbox{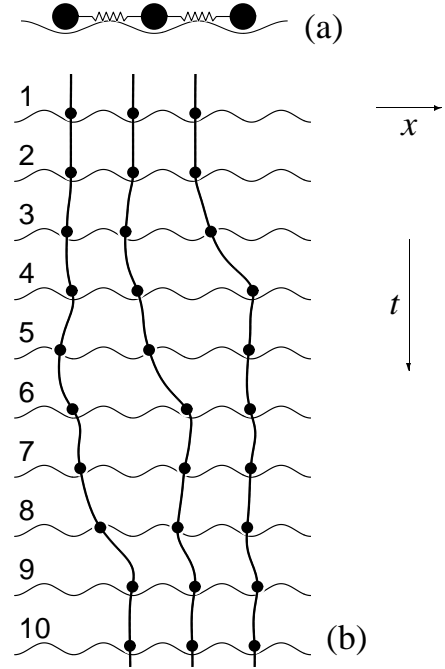}\hfill%
\vspace*{1.5mm}
\caption{(a)~Sketch of the geometry of the example engine, showing the
surface potential $\Phi(x)$ and the chain with $N = 3$ particles.
(b)~Motion of the chain sketched in (a); the position $x_i$ of the
particles as a function of time $t$ is shown. The large disks indicate
the particles' position in relation to the surface potential in the $10$
numbered snapshots in time intervals of $5 b/(2 \pi) \; \sqrt{m/\Phi_0}$.
The time $25 b/\pi \; \sqrt{m/\Phi_0}$ of a full oscillation of $a_{i,i +
\delta_i}(t)$ and hence of a single step of length $b$ to the right is
shown. The parameters are: The misfit between the minimum rest length of
the inter-particle potential and the potential period $a/b = 11/10$, the
dissipation constant $\eta = 16 \pi/(10 b) \; \sqrt{\Phi_0 m}$, the
inter-particle potential strength $k = [(2 \pi)/b]^2 \; \Phi_0$, the
driving frequency $\omega = \pi/(25 b) \; \sqrt{\Phi_0/m}$, the wave
vector $q = 1/(5b)$, the amplitude $c = 7/10$, and the peak width $s_0 =
4/10$.}
\label{fig:1Dchain}
\vspace*{-2mm}
\end{figure}

In the model system, as already mentioned, the supporting carrier is taken as
an isotropic surface, and the moving object as a chain of $N$ identical
particles on the surface. Each particle $i$ has a mass $m$ and is located at
coordinate $x_i$. For simplicity, we restrict the first part of the discussion
to translational motion in one dimension; Fig.~\ref{fig:1Dchain}(a) displays a
sketch of the model geometry for $N = 3$. The $N$ equations of motion read as
\par\vspace*{-1\baselineskip}\hspace*{0mm}
\end{multicols}
\vspace*{-6mm}
\noindent\rule{88mm}{0.5pt}\rule{0.5pt}{3mm}
\vspace*{0.5mm}
\begin{equation}\label{eq:example}
m \ddot{x}_i + \eta \dot{x}_i + \frac{\partial \Phi(x_i)}{\partial x_i} +
\sum_{\delta_i}
\frac{\partial \Psi(x_i - x_{i + \delta_i})}{\partial x_{i + \delta_i}} = 0
\qquad i = 1, \ldots, N.
\end{equation}
\vspace*{-7mm}
\hfill\rule[-2.8mm]{0.5pt}{3mm}\rule{88mm}{0.5pt}%
\vspace*{0.5mm}
\begin{multicols}{2}
\noindent
The second term in eq.~(\ref{eq:example}) describes the dissipative interaction
[friction] between the particles and the surface and is proportional to their
relative velocities with proportionality constant $\eta$. The static
interaction between the particles and the surface is represented by the
periodic potential $\Phi(x) = -\Phi_0 \; \cos(2 \pi x/b)$ with periodicity $b$.
Concerning the inter-particle interaction, we take a nearest neighbor harmonic
interaction $\Psi(x_i - x_{i + \delta_i}) = (k/2) \; [ | x_i - x_{i + \delta_i}
| - a_{i,i + \delta_i}(t)]^2$ with free equilibrium rest lengths $a_{i,i +
\delta_i}(t)$ [$i + \delta_i = i \pm 1$ denotes the nearest neightbors of
particle $i$]. The $N-1$ rest lengths depend both on the bond's position
specified by the indices $i,i + \delta_i$ and on time $t$. We demonstrate that
if energy is pumped into the system in a specific manner that provides
spatially and temporally correlated changes of the lengths $a_{i,i +
\delta_i}(t)$, the dynamical local competition between the periodicity $b$ and
the rest lengths $a_{i,i + \delta_i}(t)$ can induce a {\it directed} motion of
the chain.

Without specifing the dependence of the rest lengths $a_{i,i + \delta_i}(t)$ on
the bond's position and time, the above approach describes a whole family of
atomic scale engines. In the following we restrict ourselves to a certain
choice for $a_{i,i + \delta_i}(t)$, that results in an engine having a minimum
size of as small as $N = 3$ particles. The position and time dependent rest
lengths $a_{i,i + \delta_i}(t)$ are chosen as $a_{i,i + \delta_i}(t) = a \; [ 1
+ \alpha(q x_{i,i + \delta_i} + \omega t) ]$, resulting in a certain fixed
spatial and temporal correlation between the lengths of different bonds. Here,
$x_{i,i+1} = i b$ and $x_{i,i-1} = x_{i-1,i}$ are the relative positions of the
bonds between particles $i$ and $i \pm 1$. The length $a$, the wave vector $q$,
and the driving frequeny $\omega$ are parameters. The function $\alpha(s)$ has
a periodicity of $1$ such that $\alpha(s+1) = \alpha(s)$ for all $s$, and is
chosen as $\alpha(s) = c \sin(\pi s/s_0)$ for $0 \le s \le s_0$ and $\alpha(s)
= 0$ for $s_0 \le s \le 1$, with $s_0$ and $c$ being parameters. In
Fig.~\ref{fig:1Dchain}(b) shown are $10$ snapshots of the motion for a complete
step of length $b$ to the right for the engine sketched in
Fig.~\ref{fig:1Dchain}(a) [note that the $10^{\rm th}$ snapshot is equivalent
to the $1^{\rm st}$ as the chain moved by a length $b$ to the right]. The
direction of the motion is {\it dynamically} determined and solely given by the
bond whose rest length increases first, starting when the particles are in the
minima of the surface potential [the right one in Fig.~\ref{fig:1Dchain}(b),
cf.~\cite{Note1}]. Therefore, the motion can be easily controlled; in
particular the direction can be chosen independent of velocity, and the motion
can be stopped and restarted. The latter can be achieved with the same or with
the opposite direction. Such a control of motion is not possible with other
engines such as the ratchet systems. For the above choice of $a_{i,i +
\delta_i}(t)$ and for not too high frequencies up to a maximum frequency
$\omega_{\rm max} \approx \pi/(25 b) \; \sqrt{\Phi/m}$ [keeping the other
parameters fixed], the engine's velocity $v$ is proportional to the driving
frequency $\omega$ through $v = b \omega$, so that the maximum velocity is
approximately given by $v_{\rm max} \approx \pi/25 \; \sqrt{\Phi/m}$. For
higher driving frequencies the motion gets first irregular and finally
diffusive, loosing its directionality. The wave vector $q$ determines the
correlation between the different rest lengths. A choice of $q = 1/(5b)$ turns
out to yield a large maximum velocity and a `powerful' engine for $N = 3$. A
measure for how powerful this engine is, is determined by a simple procedure
\cite{Fisher/Kolomeisky:1999}, where the chain moves against a constant force
applied at each particle. The chain with the parameters shown in
Fig.~\ref{fig:1Dchain} is able to move against a constant force of up to
$F_{\rm max} \approx \Phi_0/b$, maintaining the velocity $b \omega$. For higher
forces, the chain first remains in its initial location, and finally moves with
the force. It is important to note that the maximum possible force depends
sensitively on the parameters, in particular on the constant $c$ and the wave
vector $q$. Another possible way to measure how powerful the engine is, is to
let the chain transport a cargo of $N'$ additional inactive particles attached
at its end with the same inter-particle potential $\Psi(x_i - x_{i +
\delta_i})$, but with a {\it constant} rest length $a$. It turns out that,
depending on the choice of parameters, a chain of $N$ particles is able to
transport up to a maximum of $N_{\rm max}' \approx N/2$ additional inactive
particles and hence up to about half its own weight on an isotropic surface.

So far, our discussion has been restricted to a linear one-dimensional system.
However, one of the main advantages of the above concept is that it can easily
be applied to other types of motion and generalized to higher dimensionality.
Let us first discuss the case
\linebreak\hspace*{1mm}\par\vspace*{-2\baselineskip}\hspace*{1mm}\par
\begin{figure}[h]
\vspace*{-3mm}
\def\epsfsize#1#2{0.76#1}
\noindent\hfill\epsfbox{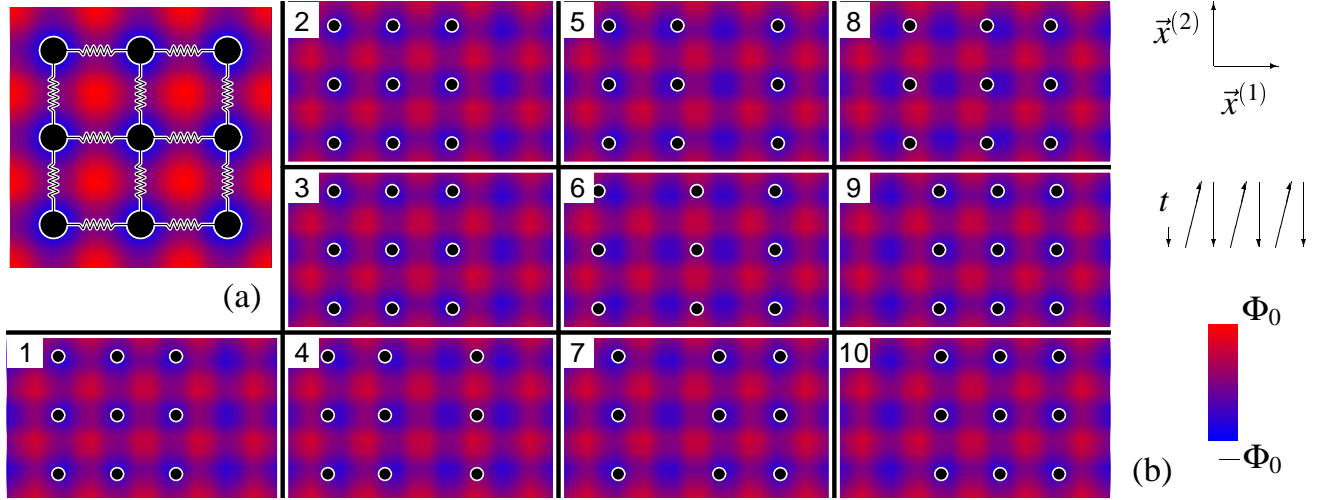}\hfill%
\caption{(color) (a)~Sketch of the 2D atomic scale `car' built by six
chains arranged in an array of $3 \times 3$ particles. The surface
potential $\Phi(\vec{x})$ is shown using the color code at the right.
(b)~Motion of the 2D atomic scale `car' sketched in (a); the black disks
indicate the particles' position in relation to the surface potential in
the $10$ numbered snapshots in time intervals of $5 b/(2 \pi) \;
\sqrt{m/\Phi_0}$. The time $25 b/\pi \; \sqrt{m/\Phi_0}$ of a full
oscillation of the respective $a_{i,i + \delta_i}(t)$ and hence of a
single step of length $b$ to the right is shown. The parameters are
identical to those in Fig.~\protect\ref{fig:1Dchain}.}
\label{fig:2Dcar}
\vspace*{-2mm}
\end{figure}

\noindent
of higher dimensionality, and consider a chain
moving on a two-dimensional surface by replacing in eq.~(\ref{eq:example}):
(a)~the 1D coordinates $x_i$ by 2D ones $\vec{x}_i$, (b)~the 1D partial
derivations $\partial/\partial x_i$ by 2D gradients $\vec{\nabla}_{\vec{x}_i}$,
(c)~the 1D surface potential $\Phi(x)$ by a 2D counterpart $\Phi(\vec{x}) =
-\Phi_0 \; \cos(\pi \; [\vec{x}^{(1)} - \vec{x}^{(2)}]/b) \; \cos(\pi \;
[\vec{x}^{(1)} + \vec{x}^{(2)}]/b)$ [$\vec{x}^{(i)}$ denotes the $i$-th
component of the vector $\vec{x}$], and (d)~the 1D inter-particle potential
$\Psi(x_i - x_{i + \delta_i})$ by the 2D counterpart $\Psi(\vec{x}_i -
\vec{x}_{i + \delta_i}) = (k/2) \; [ \big| \vec{x}_i - \vec{x}_{i + \delta_i}
\big| - a_{i,i + \delta_i}(t)]^2$ [$\big| \cdot \big|$ denotes vector length].
Using the same form for $a_{i,i + \delta_i}(t)$ as in the case of 1D, the chain
can be moved along the 2D surface. This raises the striking possibility of
building a complex atomic scale `car' by connecting, for example, six chains in
such a way that they constitute an array of $3 \times 3$ particles with $12$
bonds, see Fig.~\ref{fig:2Dcar}(a) for a sketch of the geometry. By exciting
three parallel chains coherently, this `car' can be moved both forward and
backward as well as left and right [using the six vertical or the six
horizontal bonds in Fig.~\ref{fig:2Dcar}(a)], so that it
\linebreak\hspace*{1mm}\par\vspace*{-2\baselineskip}\hspace*{1mm}\par
\end{multicols}

\begin{multicols}{2}
\noindent
can be driven {\it
freely} over the surface. As an example, in Fig.~\ref{fig:2Dcar}(b) shown are
$10$ snapshots of the motion for a complete step of length $b$ to the right
[note that the $10^{\rm th}$ snapshot is equivalent to the $1^{\rm st}$ as the
`car' moved by a length $b$ to the right].

\begin{figure}[h]
\vspace*{-3mm}
\def\epsfsize#1#2{0.76#1}
\noindent\hfill\epsfbox{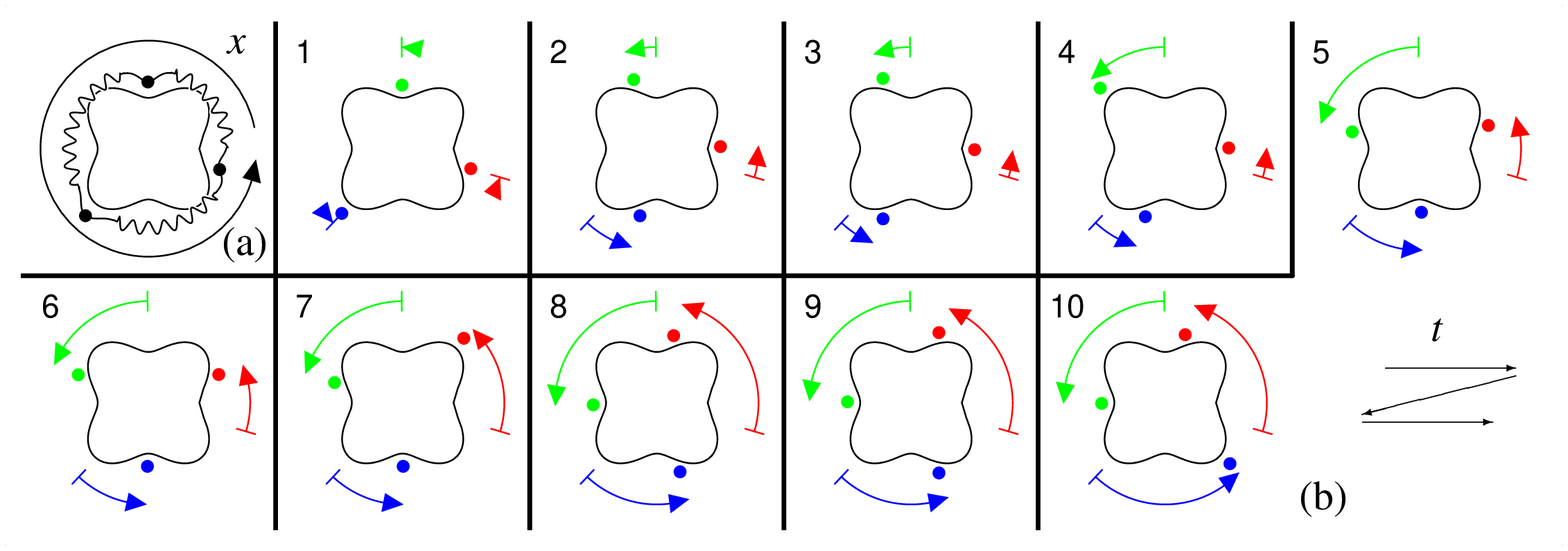}\hfill%
\caption{(color) (a)~Sketch of the atomic scale `wheel', showing the
angular surface potential $\Phi(x)$ and the ring of $N = 3$ particles on
a circle of circumfence of length $\ell = 4 b$. (b)~Motion of the atomic
scale `wheel'. The colored disks indicate the particles' position in
relation to the angular surface potential in the $10$ numbered snapshots
in time intervals of $5 b/(2 \pi) \; \sqrt{m/\Phi_0}$, where the arrows
indicate the displacement relative to the first snapshot. The time $25
b/\pi \; \sqrt{m/\Phi_0}$ of a full oscillation of $a_{i,i +
\delta_i}(t)$ and hence of a rotational step of length $b$
counterclockwise is shown [for the four-fold symmetry of the example this
is equivalent to an angle $\pi/2$]. Except $q = 1/3$ and $c = 3/10$, the
parameters are identical to those in Figs.~\protect\ref{fig:1Dchain} and
\protect\ref{fig:2Dcar}.}
\label{fig:1Dwheel}
\vspace*{-2mm}
\end{figure}

Concerning the application to other types of motion, an engine that performs
{\it rotational} motion can be achieved by treating the coordinates $x_i$ as
angular coordinates $x_i \in [0,\ell)$ on a circle of circumfence of length
$\ell = n b$ with $n$ integer, and periodic boundary conditions, see
Fig.~\ref{fig:1Dwheel}(a) for a sketch of a ring of $N = 3$ particles and $\ell
= 4 b$. This `wheel' can rotate either clockwise or counterclockwise. As an
example, in Fig.~\ref{fig:1Dwheel}(b) shown are $10$ snapshots of the rotation
for a complete step of length $b$ counterclockwise [note that the $10^{\rm th}$
snapshot is equivalent to the $1^{\rm st}$ as the `wheel' rotated by a length
$b$ counterclockwise; for the four-fold symmetry of the example, this is
equivalent to a rotation with an angle $\pi/2$ counterclockwise].

To demonstrate the robustness of our example engines and the concept in
general, we should stress a few important points. First, it is clear that the
motion of the chain is not affected by the integer part of the free rest
lengths $a_{i,i + \delta_i}(t)/b$ between the particles or `feet' that touch
the
\linebreak\hspace*{1mm}\par\vspace*{-2\baselineskip}\hspace*{1mm}\par
\end{multicols}

\begin{multicols}{2}
\noindent
surface. In the above examples in Figs.~\ref{fig:1Dchain} and
\ref{fig:2Dcar}, we have chosen the rest lengths $a_{i,i + \delta_i}(t)$ to
oscillate between $11 b/10$ and $187 b/100$ [a relative change by $187/110$],
but the same motion occurs for $a_{i,i + \delta_i}(t)$ oscillating between $21
b/10$ and $287 b/100$ [a relative change by $287/210$], between $31 b/10$ and
$387 b/10$ [a relative change by $387/310$], and so on. Hence, the relative
change can be made arbitrarily small. A second possibility to reduce the
required stretching of the bond needed to cause a directed motion is to divide
the $N - 1$ bonds into $g$ groups [$N$ bonds in the case of rotational motion],
each containing $(N-1)/g$ bonds [$N/g$ bonds in the case of rotational motion],
by setting $x_{i,i+1} = {\rm int}(i/g) \; b$ [${\rm int}(x)$ denotes the
largest integer $n$ with $n \le x$]. For instance, for the parameters used for
Figs.~\ref{fig:1Dchain} and \ref{fig:2Dcar}, but with $N = 10$ particles [i.e.\
$9$ bonds] and $g = 3$ groups, each containg $3$ bonds, it is sufficient that
the free rest lengths $a_{i,i + \delta_i}(t)$ oscillate between $b$ and $142
b/100$ [instead of $11 b/10$ and $187 b/100$] to cause a directed translational
motion with the same velocity $b \omega$.
\linebreak\hspace*{1mm}\par\vspace*{-2\baselineskip}\hspace*{1mm}\par

One important question is that of the influence of the shape of the
`excitation' $\alpha(s)$, which is sinusodial in our case. As long as the
`excitation' is approximately trapezoidal, a proper choice of driving frequency
$\omega$ and wave vector $q$ results in a directed motion. Since
eq.~(\ref{eq:example}) is deterministic, another important question is that of
the influence of noise. As long as the thermal energy $k_{\rm B} T$ is much
smaller than the energy scale $\Phi_0$, the engines motion is hardly influenced
by noise. If the fluctuations become larger, the motion gets erratic. Finally,
for $k_{\rm B} T \gg \Phi_0$, the motion looses completly its directionality
and becomes diffusive.

In comparison with other existing approaches to translational atomic scale
engines, e.g.\ the ratchet systems mentioned earlier, or other models that use
a spring with a time-dependent length in combination with static/dynamic
friction \cite{Mogilner/Mangel/Baskin:1998} or with a spatial asymmetric
potential \cite{Stratopoulos/Dialynas/Tsironis:1999}, the current approach has
a number of advantages. In particular, our atomic scale engines can operate as
microscopic `shuttles' or `trucks' that transport a cargo, a property that is
absent in all other proposed schemes. In addition, the engines proposed here
are not bound to a one-dimensional track and can be applied to different
geometries and higher dimensionality.

As a conclusion, we would like to emphasize that the choice of a chain on a
surface as the example engine has been motivated solely by the simplicity of
the model rather than by experimental requirements. One important feature of
the concept introduced here is that it does not impose any specific length or
time scales. This means that it is applicable not only for an engine on an
atomic scale, but also for mesoscopic or even macroscopic sizes. Hence, the
basic concept of competing lengths is more general than presented here and
might be applicable in a wide range of situations. A possible realization on an
atomic or mesoscopic scale is by building the moving object using nanosize
clusters [the `particles'] and photochromic molecules [the `bonds']. The time
dependence of the rest lengths, which is a crucial part of the model, can be
provided by individually controlling the `bonds' by light induced
conformational changes of the chromophors. Using different chromophores which
respond to different wavelengths should allow to specifically excite `bonds' at
chosen locations at given times. In this proposed realization the surface
corrugation can be prepared by nanolithography. We believe therefore that the
current concept is simple and robust enough so that it can be realized in
actual experiments using already existing techniques. Such an atomic scale
engine, when realized experimentally, will be a breakthrough in the field of
nanotechnology in providing new ways to manipulate molecules and clusters.

Financial support from the Israel Science Foundation, the German Israeli
Foundation, and DIP and SISITOMAS grants is gratefully acknowledged. M.P.\
gratefully acknowledges the Alexander von Humboldt Foundation (Feodor Lynen
program) for financial support.

\end{multicols}

\end{document}